\documentclass{PoS}

\usepackage{graphicx}
\usepackage[utf8]{inputenc}
\usepackage[english]{babel}
\usepackage{amsmath,amssymb,color,amscd,subfigure}

\usepackage{setspace}
\providecommand*{\ord}{\ensuremath{\text{O}}}

\providecommand*{\De}{\ensuremath{\mathcal{D}}}
\providecommand*{\de}{\ensuremath{\rm{d}}}

\title{A simple method to optimize HMC performance}

\ShortTitle{Optimize HMC performance}

\author{ \speaker{A. Bussone}$^a$, M. Della Morte$^a$, V. Drach$^b$, 
M. Hansen$^a$,  A. Hietanen$\,^a$, J. Rantaharju$^a$,
  C. Pica$\,^a$ \\
  \llap{$^a$}{CP$^3$-Origins, University of Southern Denmark, Campusvej 55, DK-5230 Odense M, Denmark\\}
  \llap{$^b$}{CERN, Physics Department, 1211 Geneva 23, Switzerland \\}
E-mail:\email{bussone@cp3-origins.net}
}

\abstract{%
{\it Preprint: CP3-Origins-2016-038 DNRF90URL}\\
\\
We present a practical strategy to optimize a set of Hybrid Monte Carlo parameters in simulations of QCD and QCD-like theories. We specialize to the case of mass-preconditioning, with multiple time-step Omelyan integrators. Starting from properties of the shadow Hamiltonian we show how the optimal setup for the integrator can be chosen once the forces and their variances are measured, assuming that those only depend on the mass-preconditioning parameter.
}

\FullConference{The 34th International Symposium on Lattice Field Theory\\
                 24 - 30 July  2016\\
                 Southampton, United Kingdom}

\begin{document}

\section{Introduction and basic definitions}

Modern HMC algorithms require tuning of multiple parameters for an ef\mbox{}ficient
generation of configurations. The optimization of the parameters is a
complicated task but has been achieved in multiple ways in QCD \cite{Urbach:2005ji}. Here we
present a general strategy, based on the existence of a shadow Hamiltonian,
that can also be extended to strongly interacting BSM (Beyond Standard Model)
theories.\\
In the following we specialize ourself to the case of the Omelyan integrator \cite{Omelyan:2003} with
$\alpha = 1/6$. We consider an $\mathbf{SU}(2)$ gauge group with a doublet of unimproved Wilson fermions
in the fundamental representation. For completeness the bare parameters of the simulation read $\beta = 2.2$ and $m_0 = -0.72$. For comparison the 
critical mass is estimated to be $m_{\rm cr}\simeq -0.77(1)$\footnote{We recall that theory is breaking chiral symmetry spontaneously and the current investigation show the theory is QCD-like.}.

\subsection{Shadow Hamiltonian}

To each symplectic integrator corresponds an exactly conserved shadow Hamiltonian. In order to
introduce it we begin by writing the evolution operator as
\begin{align*}
\exp\left(\tau\frac{\de}{\de t}\right)\star = \exp(\{H,\star\}) \equiv \exp(\tau D_H)\star.
\end{align*}
where $\tau$ is a fictitious time parameter, $\{\cdot, \cdot\}$ the Poisson bracket, $H = T + S$ is the Hamiltonian of
the system, $T$ the kinetic part for the conjugate momenta, $S$ the action we want to simulate, and
$D_H \equiv D_{T} + D_{S}$ the associated operators. The Omelyan integrator is then given by the following
evolution operator
\begin{align*}
\exp\left[\tau \alpha D_S\right] \exp\left[\frac{\tau}{2} D_T\right] \exp\left[\tau (1-\alpha) D_S\right] \exp\left[\frac{\tau}{2} D_T\right] \exp\left[\tau \alpha D_S\right],
\end{align*}
with $\alpha$ being a free parameter. By using the Baker-Campbell-Haussdorf\mbox{}f (BCH) formula one obtains that
the conserved shadow Hamiltonian $\widetilde{H}$ is related to the target one $H$ to be simulated, by
\begin{align*}
\widetilde{H} = H + \tau^2\left\{\frac{6\alpha^2-6\alpha+1}{12} D_S[D_S(T)] + \frac{1-6\alpha}{24} D_T[D_S(T)] \right\} + \ord(\tau^4).
\end{align*}
By setting $\alpha = 1/6$ the second term vanishes and what remains is dependent on $D_S[D_S (T)] = \{S, \{S, T\}\}$,
which as we will see in the following is directly related to the forces entering in the molecular dynamics simulations.

\subsection{Mass preconditioning \& multi time-scale}

A way to reduce the fluctuations of the force is to employ \emph{mass preconditioning} of the quark
determinant \cite{Hasenbusch:2002ai}. The definitions for the massive and hermitian Dirac operators are the following
\begin{align*}
D_m = D+m,\quad Q=\gamma_5 D_m.
\end{align*}
In the presence of mass preconditioning the probability distribution for the generation of configurations splits in three parts
\begin{align*}
P_S \propto \int \De[\phi_i,\phi_1^\dagger, \phi_2, \phi_2^\dagger] \exp\bigg( -\underbrace{S[U]}_{\rm Gauge} - \underbrace{\phi_1^\dagger ( DD^\dagger +\mu^2)^{-1}\phi_1}_{\rm HMC} -\underbrace{\phi_2^\dagger \left(\frac{Q^2}{DD^\dagger +\mu^2}\right)^{-1}\phi_2}_{\rm Hasenbusch} \bigg).
\end{align*}
Hence we are now dealing with three forces: Gauge, HMC, Hasenbusch.
A further acceleration can be achieved by considering multiple time-step integrators \cite{Urbach:2005ji}, which
consists of taking dif\mbox{}ferent integration step sizes for the dif\mbox{}ferent forces. We assume that in the
outermost level there is the evolution for $S_1$ with time step $\delta\tau = \tau_f /n$, in the middle the integrator for
$S_2$ with $m$ steps and the innermost is for $S_3$ with $k$ steps.\\
The shadow Hamiltonian associated to the Omelyan integrator with three time-scales and mass
preconditioning is a quite lengthy expression but by setting the parameter $\alpha = 1/6$ it is given by
\begin{align*}
\widetilde{H} = H + \frac{\delta\tau^2}{72}\left[ \{S_1, \{S_1, T\}\} + \frac{1}{4m^2} \{S_2, \{S_2, T\}\}  +\frac{1}{16m^2k^2} \{S_3, \{S_3, T\}\}  \right] + \ord(\delta\tau^4).
\end{align*}
We use the conventions adopted in \cite{Kennedy:2012gk, Kennedy:2007cb} and the above formula reduces to
\begin{align}
\label{eq:var}
\widetilde{H} &= H + \frac{\delta\tau^2}{72} \sum_{x,\mu,a}  \left[ T_{R,1}\left(F_1^{a\mu}(x)\right)^2  + T_{R,2}\frac{\left(F_2^{a\mu}(x)\right)^2}{4m^2} + T_{R,3}\frac{\left(F_3^{a\mu}(x)\right)^2}{16m^2k^2}  \right] + \ord(\delta\tau^4)\\
&\equiv H + \frac{\delta\tau^2}{72} \sum_{x,\mu,a}  \left( |\mathcal{F}_1|^2 + \frac{|\mathcal{F}_2|^2}{4m^2}  + \frac{|\mathcal{F}_3|^2}{16m^2k^2}  \right) + \ord(\delta\tau^4) \equiv H + \delta H + \ord(\delta\tau^4),
\nonumber
\end{align}
that is a function only of the forces used during the simulations.
We immediately see that the shadow Hamiltonian is related to the dif\mbox{}ferent parts of the force
weighted by the corresponding normalization for the generators. Already at this point one can
see what drove our assignments for the dif\mbox{}ferent levels. We want to suppress the contribution of the \emph{bigger}
force (the gauge one) and hence that will go in the innermost level, followed by the HMC force and finally the Hasenbusch one at the outermost level.

\section{Benchmarks in small volumes}

\begin{figure}[h!t]
\begin{center}
\subfigure[\emph{Scaling of $|\Delta H|$ and $|\Delta(\delta H) + \Delta H|$ with $\delta \tau$.}]{\includegraphics[scale=0.5]{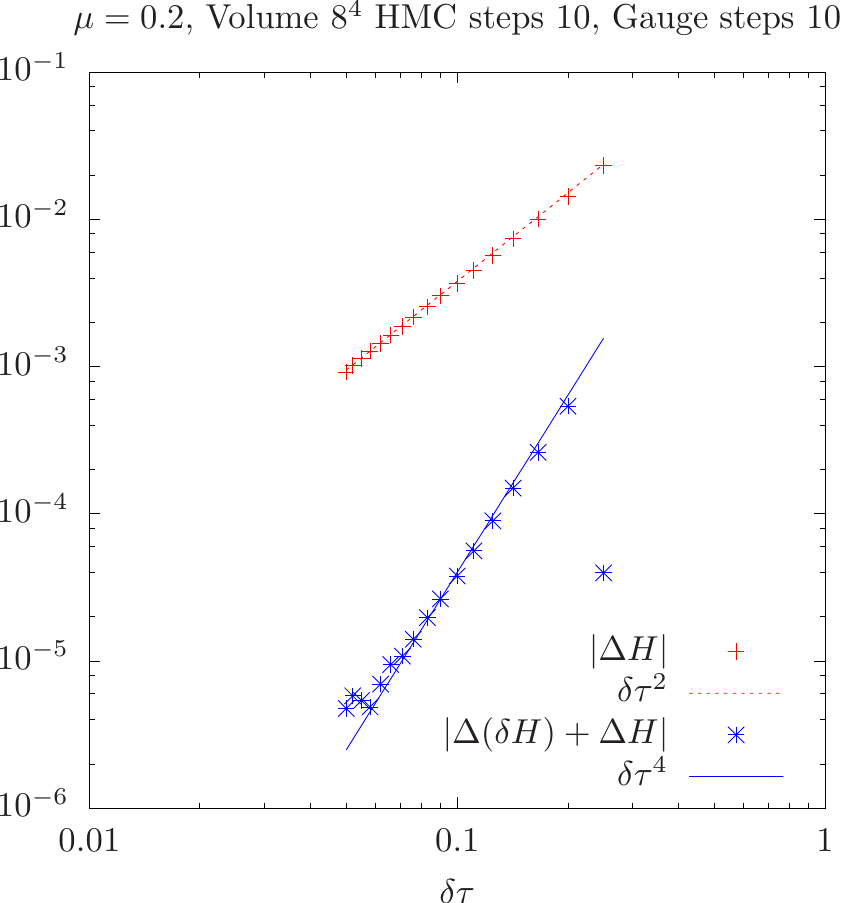}}
\subfigure[\emph{$\Delta H$ history along a trajectory with $n=10$.}]{\includegraphics[scale=0.5]{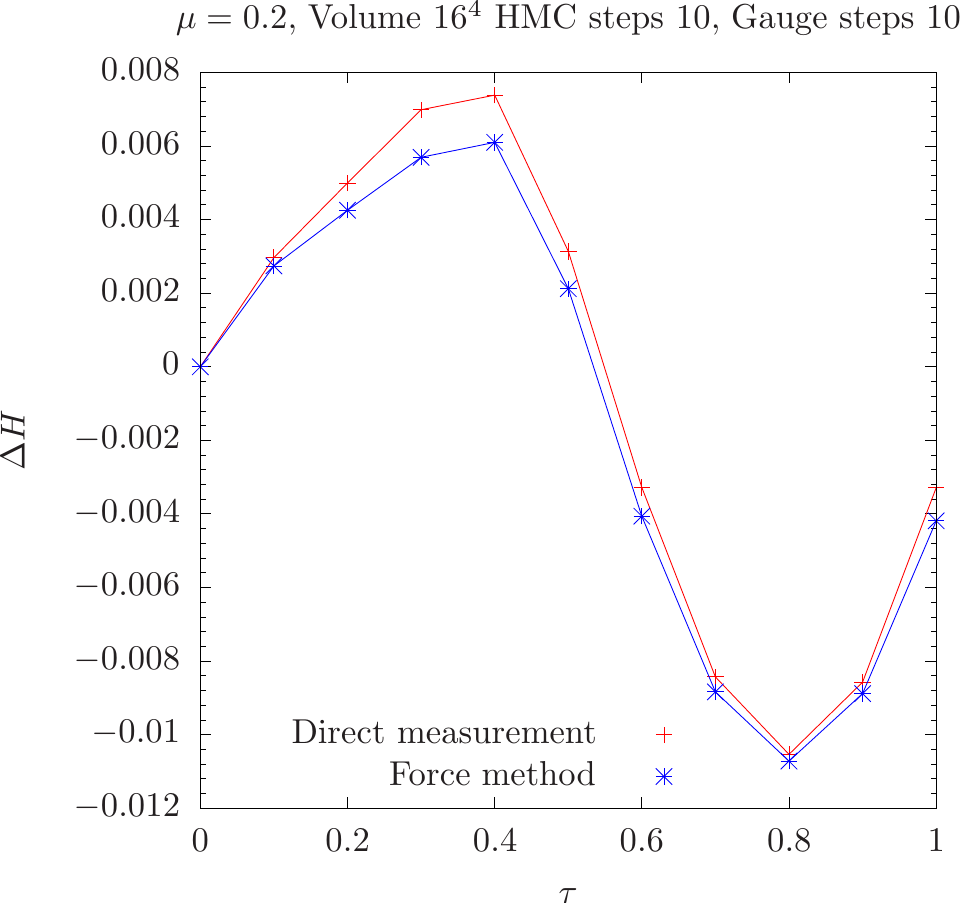}\label{fig:delta_h_10}}
\subfigure[\emph{$\Delta H$ history along a trajectory with $n=20$.}]{\includegraphics[scale=0.5]{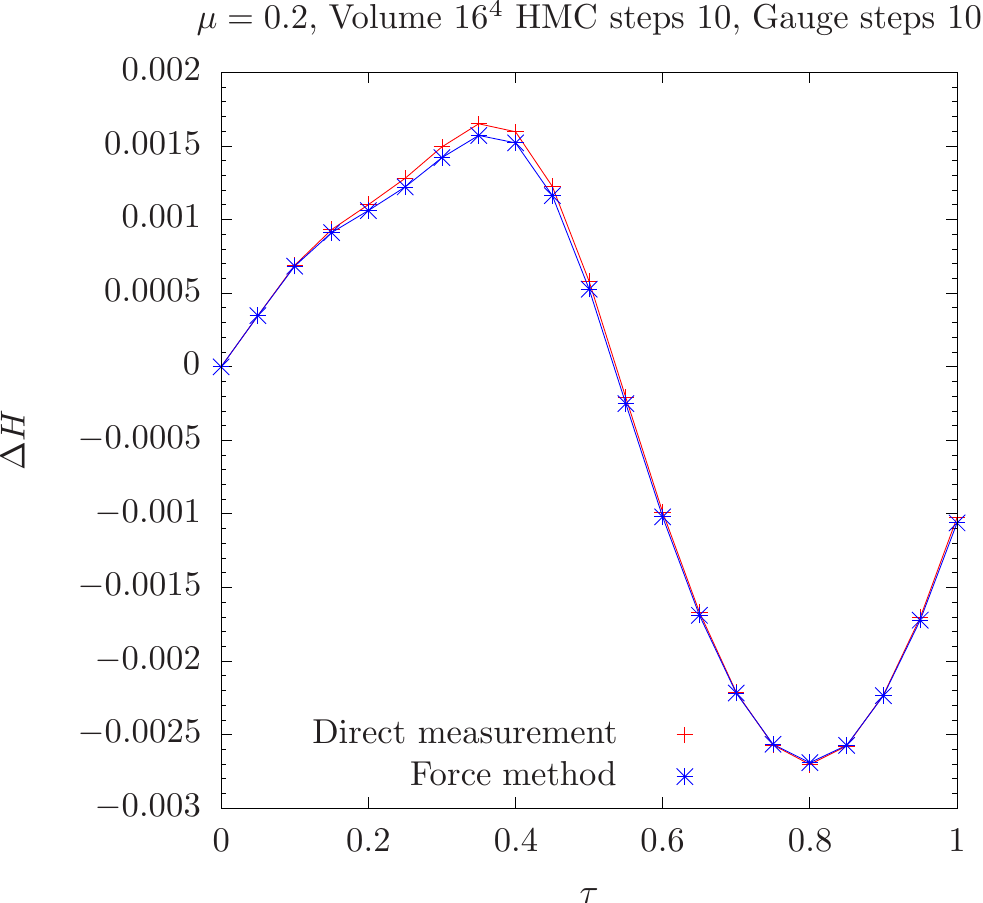}\label{fig:delta_h_20}}
\caption{\emph{Benchmarks for the Poisson brackets measurements.}}
\label{fig:benchmarks}
\end{center}
\end{figure}

In order to test the measurements of the Poisson brackets we check the scaling with $\delta\tau$ of $|\Delta H|\propto\delta\tau^2$ and $|\Delta(\delta H) + \Delta H|\propto\delta\tau^4$,
indeed, since $\widetilde{H}$ is conserved along the trajectory, we have
\begin{align}
\Delta \widetilde{H} = 0 = \Delta H + \Delta(\delta H) + \ord(\delta\tau^4) \Longrightarrow \Delta H = - \Delta(\delta H) + \ord(\delta\tau^4).
\label{eq:delta}
\end{align}
Another test is to measure directly $\Delta H$ along the trajectory and compare it with the one built from
the knowledge of the forces, eq.~\eqref{eq:delta}. 
We run one trajectory from a thermalized configuration with the following set-up for the levels of
integration
\begin{itemize}
\setlength\itemsep{-0.5em}
\item level 0: Hasenbusch, $n = 4,5,\dots,20$,
\item level 1: HMC, $m = 10$,
\item level 2: Gauge, $k = 10$.
\end{itemize}
The results for $8^4$ and $16^4$ volumes are showed in Fig.~\ref{fig:benchmarks}.
It is worth to notice that the minimum of $\Delta H$ scales as predicted and when the minimum (or a
maximum) is attained, then $\Delta H$ cannot grow due to the existence of the shadow Hamiltonian, 
and it is well understood in terms of the various underlying force contributions,
see Figs.~\ref{fig:delta_h_10}, \ref{fig:delta_h_20}.

\section{Cost of a simulation and its minimization}

\begin{figure}[t!]
\begin{center}
\includegraphics[scale=0.7]{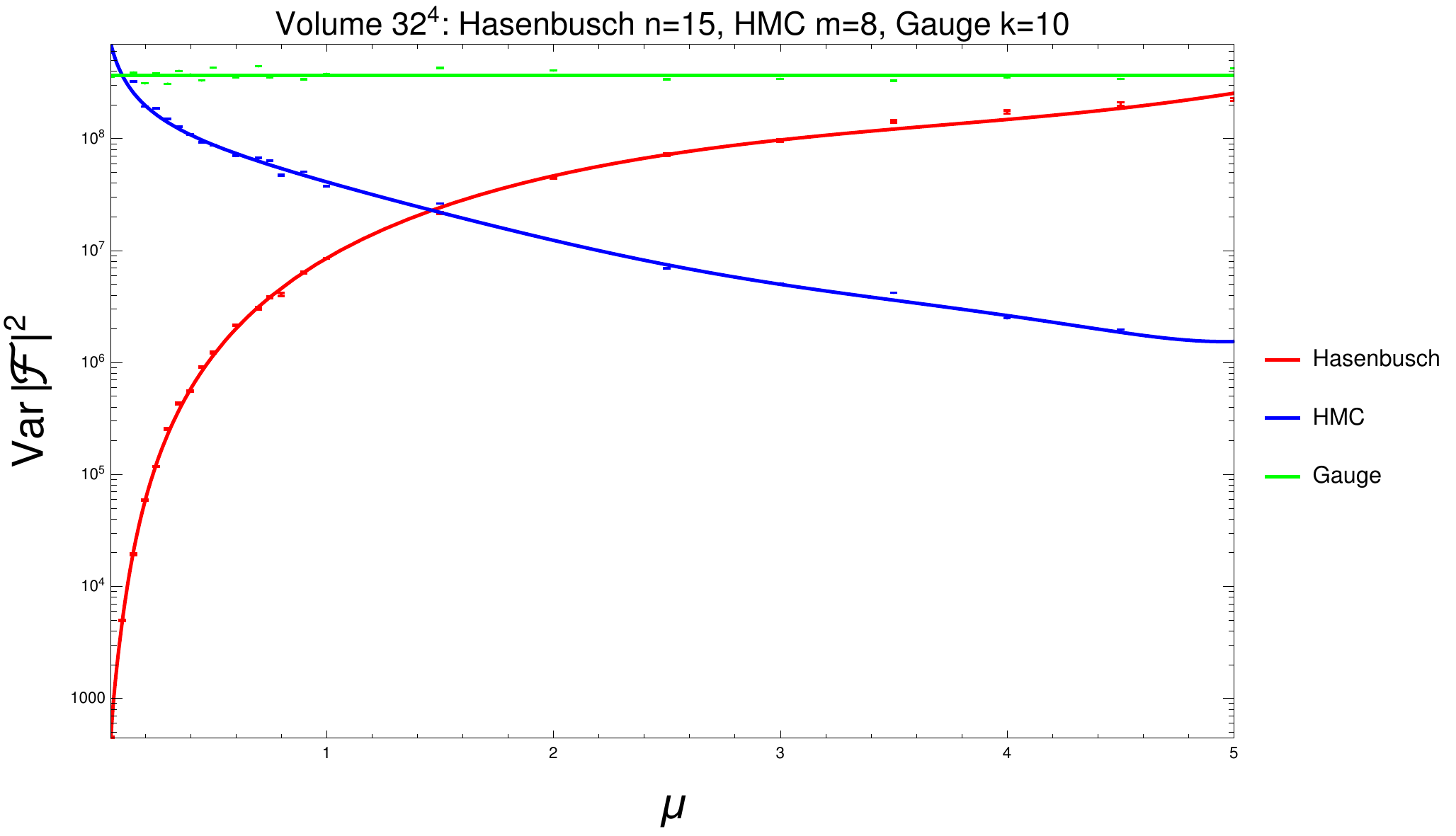}
\caption{\emph{${\rm Var}(|\mathcal{F}_i|^2)$ as a function of $\mu$ for the different forces.}}
\label{fig:var_mu_dep}
\end{center}
\end{figure}

\begin{figure}[h!t]
\begin{center}
\subfigure[\emph{$\#{\rm MVM}_1(\mu)$ data and their fit.}]{\includegraphics[scale=0.4]{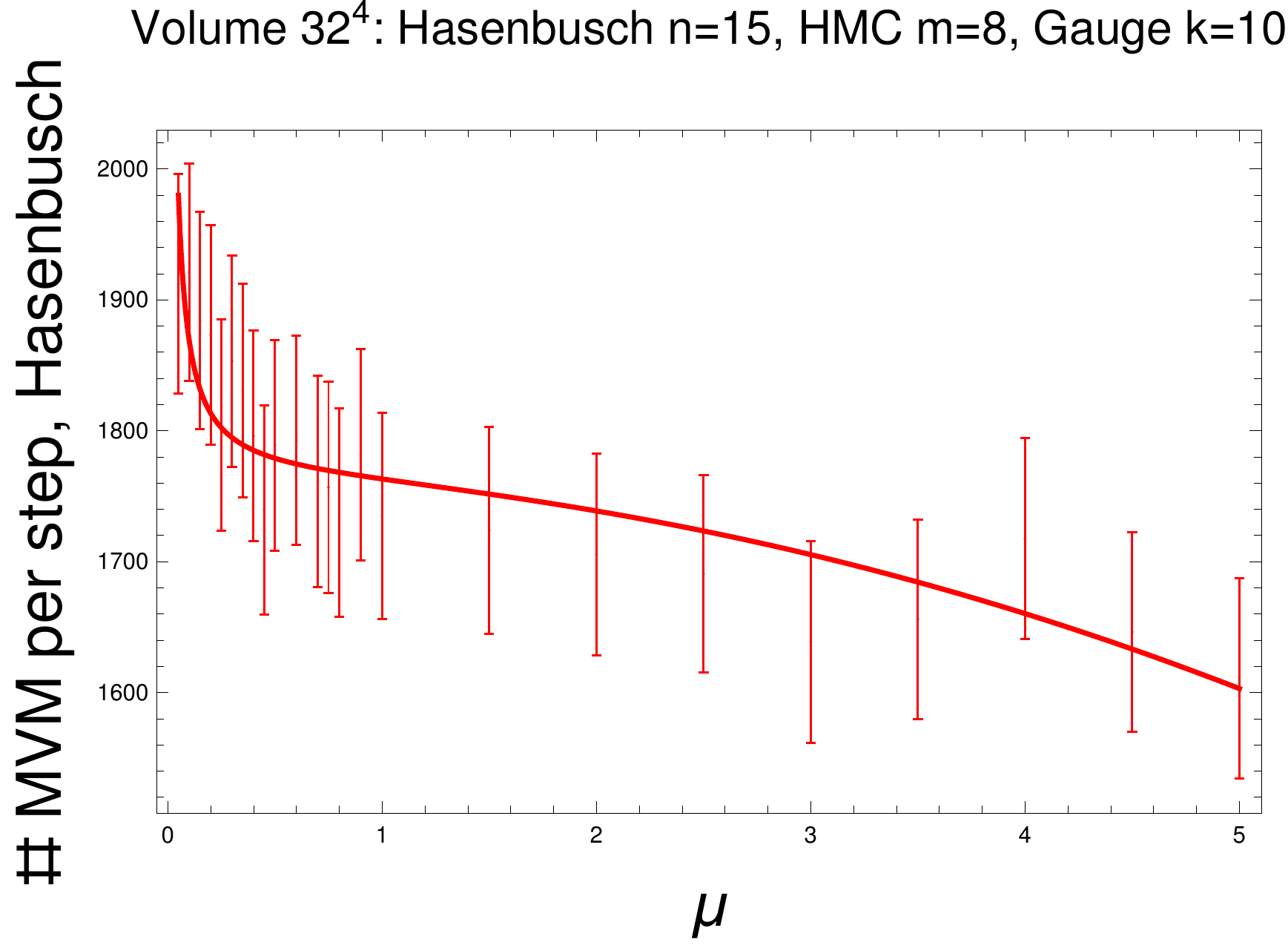}\label{fig:mvm_hase}}
\hspace{0.5cm}
\subfigure[\emph{$\#{\rm MVM}_2(\mu)$ data and their fit.}]{\includegraphics[scale=0.4]{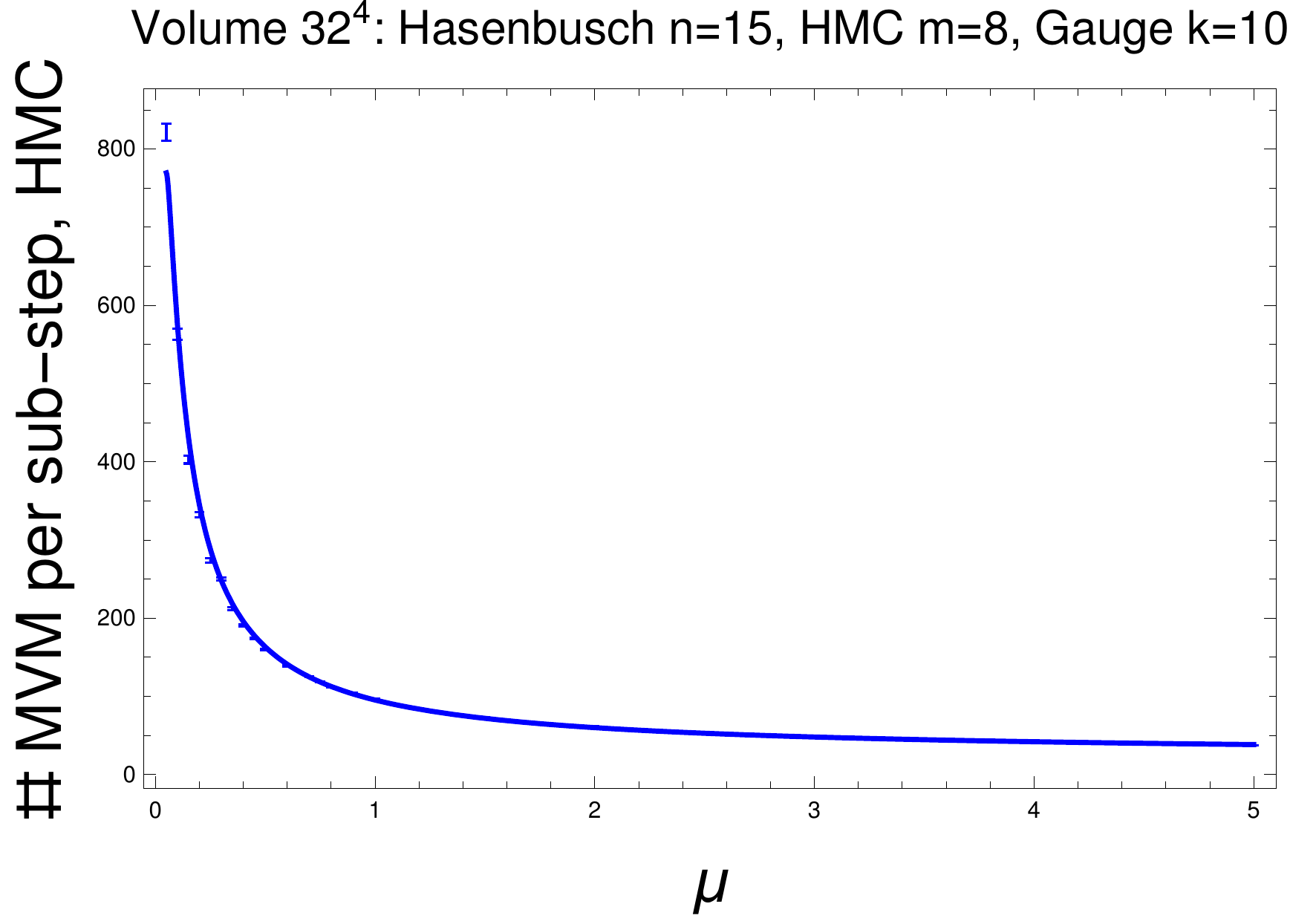}\label{fig:mvm_hmc}}
\caption{\emph{Fitted quantities for the prediction of the acceptance and the cost function.}}
\label{fig:mvm_mu_dep}
\end{center}
\end{figure}

Although the cost of a simulation is not unique we define it as
\begin{align*}
{\rm Cost} = \frac{\# {\rm MVM}}{P_{\rm acc}}.
\end{align*}
The number of Matrix-Vector-Multiplications (\# MVM) is machine independent\footnote{The
gauge part contributes for a maximum of $\sim$5\% of the cost, hence is negligible. We also took into account the
gauge part as a check and it does not af\mbox{}fect the results of this work.} . Furthermore we
neglect the autocorrelation since that conceivably has a mild dependence on $\mu$ and therefore should be mostly contribute as an overall factor to the cost.\\
We link the acceptance $P_{\rm acc}$ to $\Delta H$ through the Creutz formula \cite{Gupta:1990ka}
\begin{align*}
P_{\rm acc} (\Delta H) = {\rm erfc}\left(\sqrt{{\rm Var}(\Delta H)/8}\right),
\end{align*}
and the connection between the variances of $\Delta H$ and $\delta H$ is given by \cite{Clark:2008gh}
\begin{align*}
{\rm Var} (\Delta H) \simeq 2{\rm Var}(\delta H).
\end{align*}
We want to optimize the choice of parameters $\mu, n, m, k$ while keeping the integrator, the solver and the number of Hasenbusch splittings fixed.
The variance of $\delta H$ can be written by using eq.~\eqref{eq:var}, and neglecting the covariances\footnote{We checked that those are indeed of negligible size.}
\begin{align}
\label{eq:var_delta_h}
{\rm Var}(\delta H) \simeq \frac{\delta\tau^4}{(72^2)}\left[{\rm Var}(|\mathcal{F}_1|^2)(\mu) + \frac{{\rm Var}(|\mathcal{F}_2|^2)(\mu)}{(4m^2)^2} + \frac{{\rm Var}(|\mathcal{F}_3|^2)(\mu)}{(16m^2k^2)^2}\right].
\end{align}
The total average number of MVM is given in terms of the averages a each level by
\begin{align}
\label{eq:mvm}
\# {\rm MVM} = (2n+1) \# {\rm MVM}_1(\mu) +  2n(2m+1) \# {\rm MVM}_2(\mu).
\end{align}
The idea is to assume that ${\rm Var} (\delta H)$ and \# MVM depend explicitly upon $n, m$ and $k$ as in
eqs.~(\ref{eq:var_delta_h}, \ref{eq:mvm}) and the dependence on $\mu$ of ${\rm Var}(|\mathcal{F}_i|^2)$ at fixed $m_0$ is the only quantity to be modeled, see Figs.~\ref{fig:var_mu_dep}, \ref{fig:mvm_mu_dep}.\\
In Fig.~\ref{fig:var_mu_dep} we show the variances for the dif\mbox{}ferent forces and their resulting fits. We can identify two
dif\mbox{}ferent regions: a \emph{strong} dependence for small $\mu$ and a \emph{weak} dependence for large $\mu$. In the weak
dependence region we have an inverted hierarchy with respect to what was our choice. In Figs.~\ref{fig:mvm_hase}, \ref{fig:mvm_hmc} we show the number of MVMs per step and sub-step and their fits.


\begin{figure}[h!t]
\begin{center}
\subfigure[\emph{${\rm Cost / Cost_{min}}$ with $m=3$ as  function of $\mu$ and $n$.}]{\includegraphics[scale=0.5]{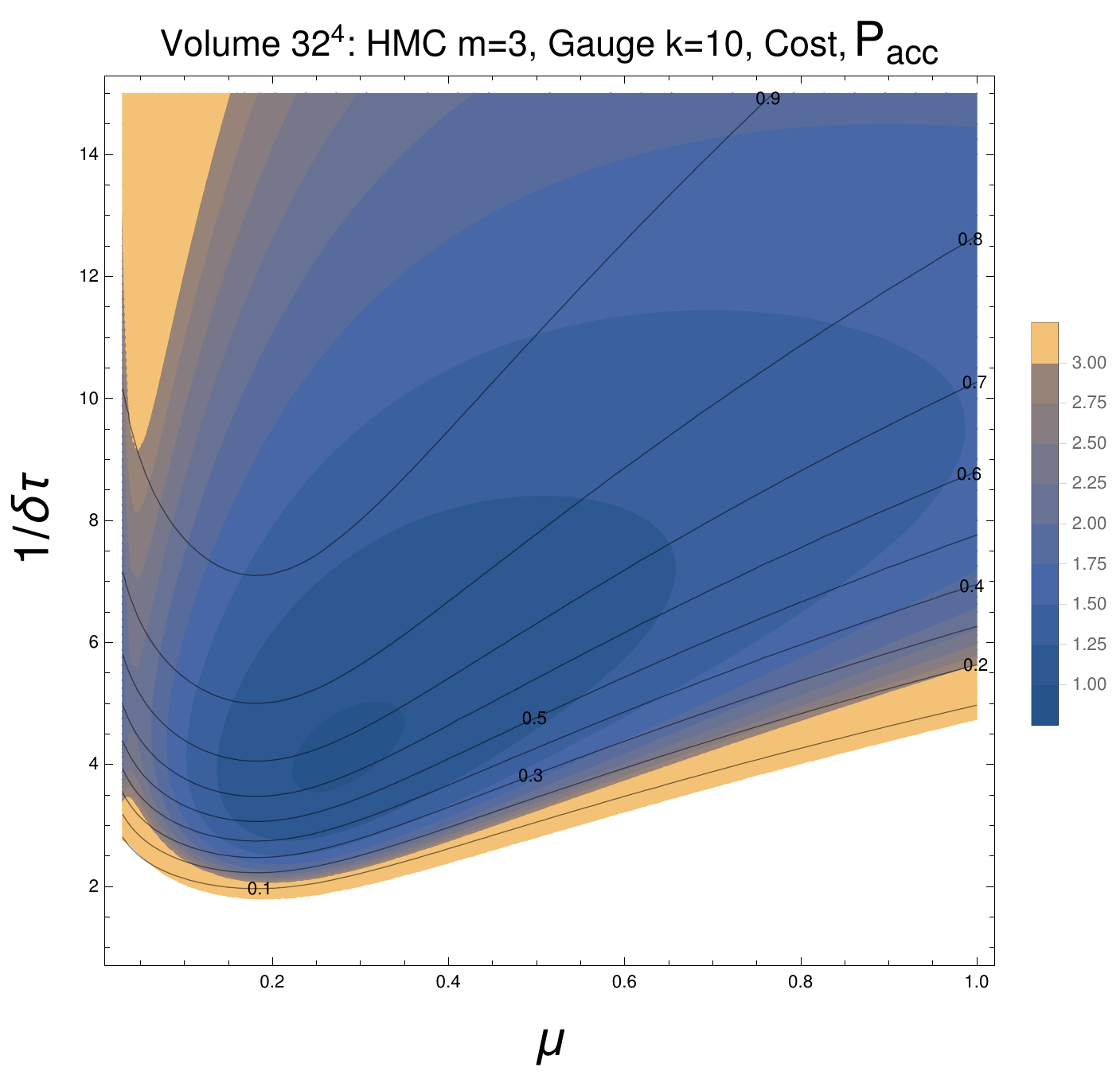}\label{fig:cost_min_hase}}
\hspace{0.5cm}
\subfigure[\emph{${\rm Cost / Cost_{min}}$ with $n=5$ as  function of $\mu$ and $n$.}]{\includegraphics[scale=0.5]{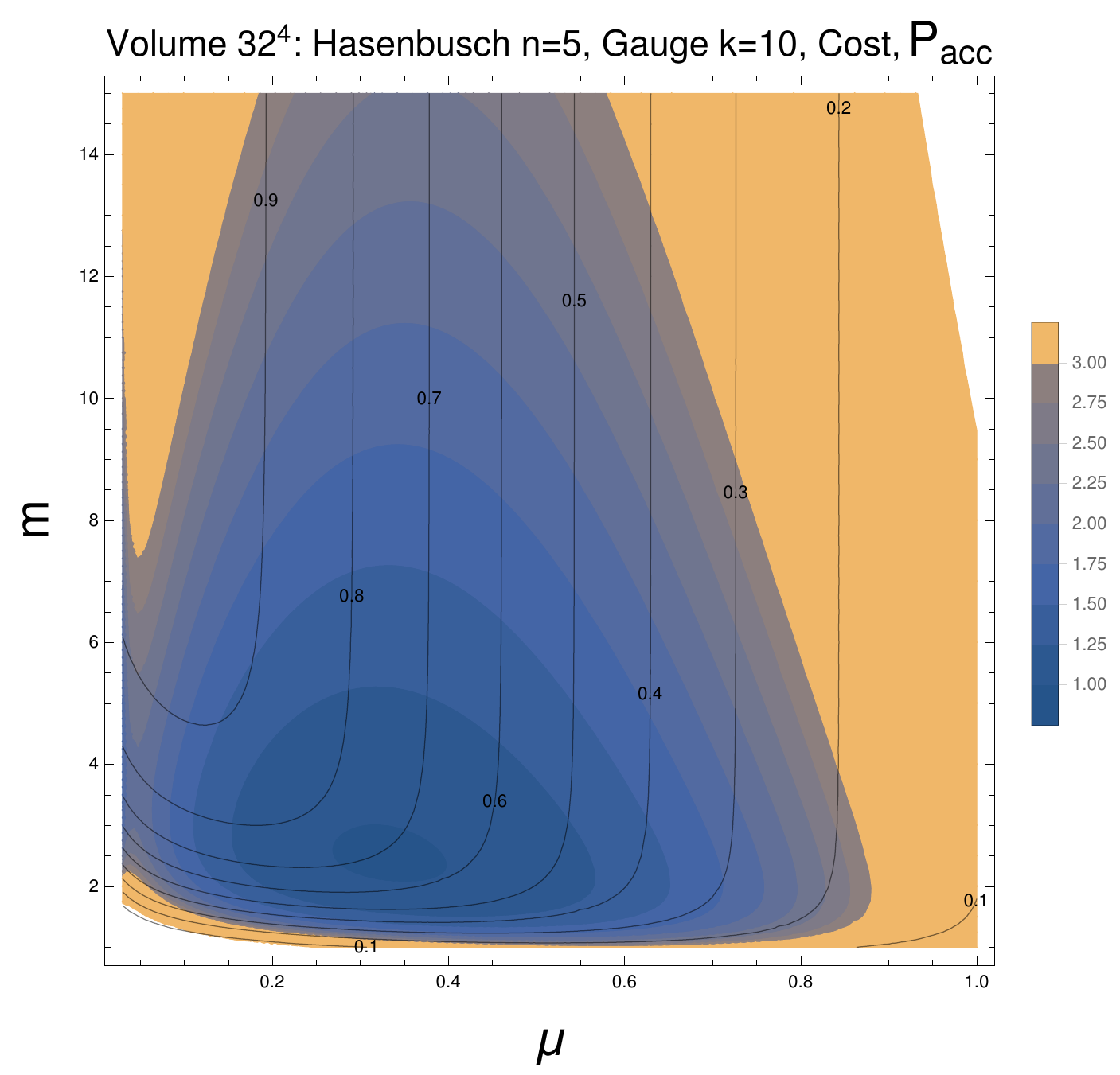}\label{fig:cost_min_hmc}}
\caption{\emph{${\rm Cost / Cost_{min}}$ around the minimum $(n,m,k)\simeq(5, 3, 0.3)$.}}
\label{fig:cost_min}
\end{center}
\end{figure}

We can now build the cost as a function of $n, m, k$ and $\mu$. For simplicity we fix $k = 10$ and in
order to find the minimum in the other parameters we require $P_{\rm acc} \gtrsim 70\%$\footnote{This requirement is needed for Creutz formula
to hold true.}
. With this set-up we
found the minimum, ${\rm Cost_{min}}$, to be at $(n, m, \mu) \simeq (5, 3, 0.3)$. In Fig.~\ref{fig:cost_min} we show the cost, normalized to
the minimum, and the acceptance in the plane $(\mu, n)$ and $(\mu, m)$, for definiteness $\beta=2.2$, $m_0=-0.72$, $V=32^4$. In Fig.~\ref{fig:cost_min_hase} we fix $m = 3$, and one can
see that the minimum is close to the boundary $P_{\rm acc}\sim 70\%$. By taking the cost of a simulation to
be $1 < {\rm Cost} / {\rm Cost_{min}} < 1.25$ we see that the minimum is quite broad $0.2\lesssim\mu\lesssim 0.6$ and $4 \lesssim n \lesssim 8$.
Same conclusions can be drawn for the Fig.~\ref{fig:cost_min_hmc}.

\subsection*{Comparison with simulation}

\begin{figure}[h!t]
\begin{center}
\subfigure[\emph{Comparison of the acceptance at the predicted minimum and a simulation.}]{\includegraphics[scale=0.45]{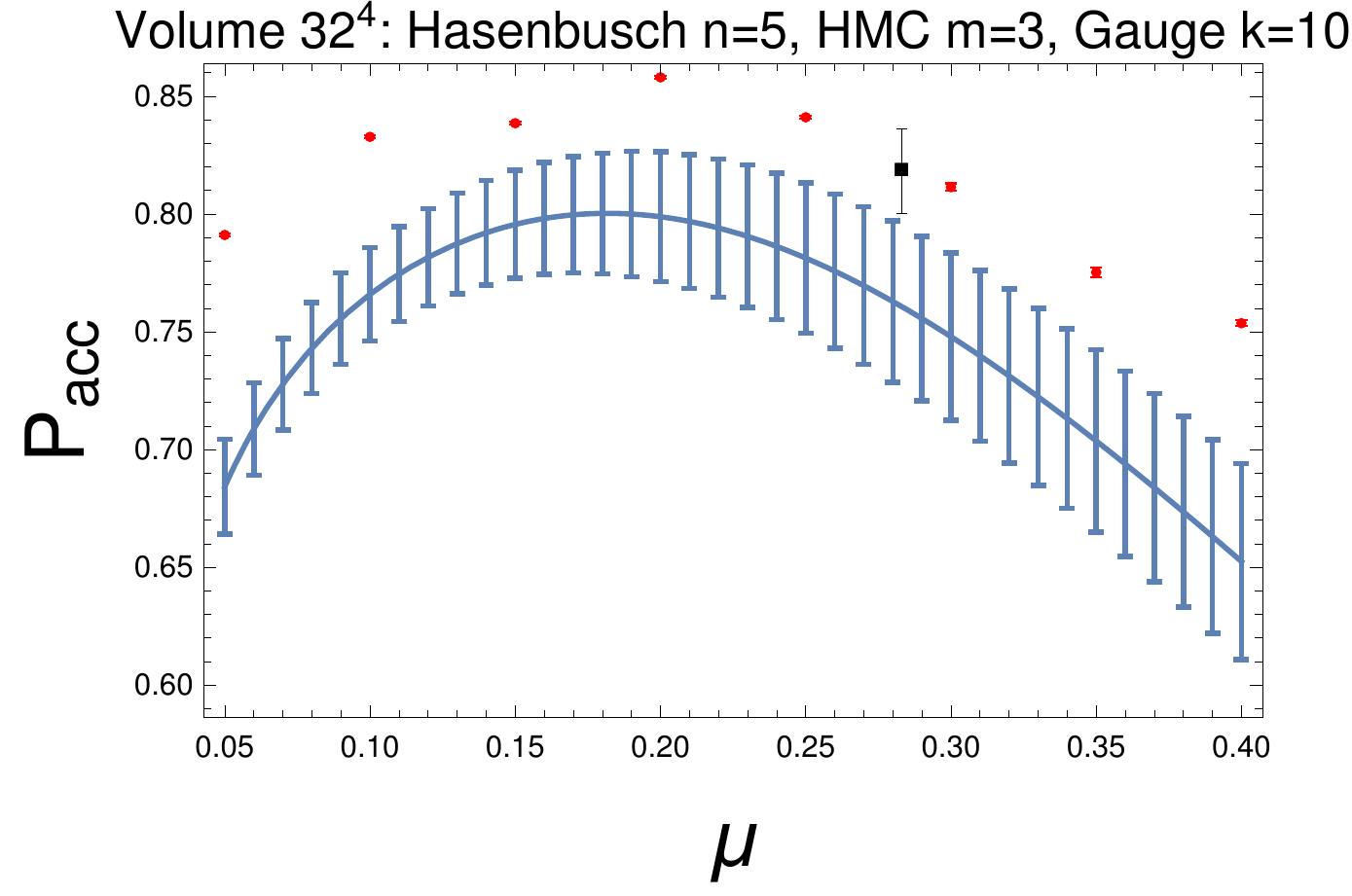}\label{fig:sim_min_hase}}
\hspace{0.5cm}
\subfigure[\emph{Comparison of the cost at the predicted minimum and a simulation.}]{\includegraphics[scale=0.45]{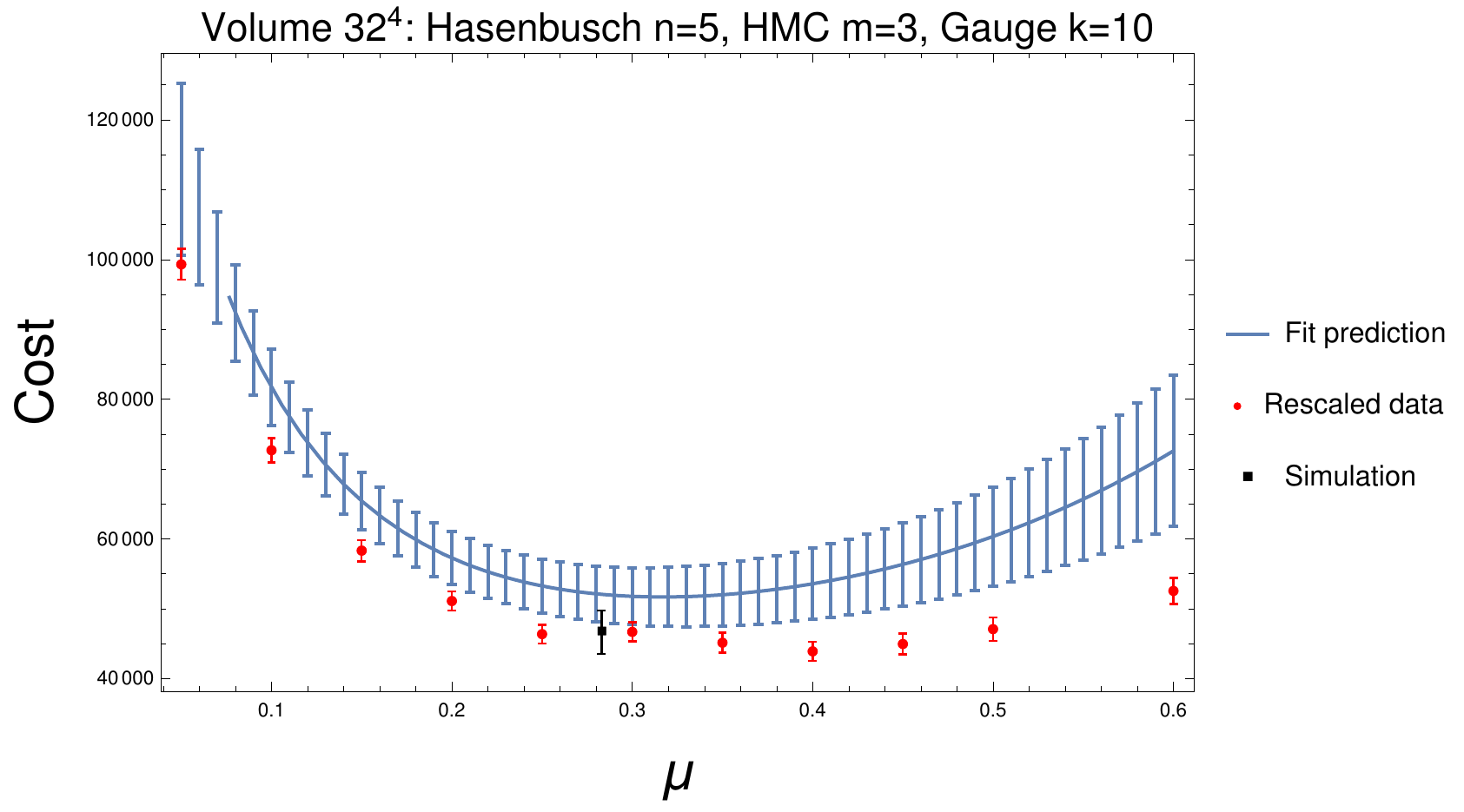}\label{fig:sim_min_hmc}}
\caption{\emph{Acceptance and cost at the minimum, comparison with simulation.}}
\label{fig:sim_min}
\end{center}
\end{figure}

We have run a simulation around the minimum to test our assumptions so far. The results are
shown in Fig.~\ref{fig:sim_min}, the blue line displays the prediction with the procedure described above, the circle
points are the directly computed raw data (no fit in $\mu$ was performed) and the square point comes from
the simulation. The results agree with both procedures within 10\% that we consider
satisfactory for the approach used and our goals.

\section{Conclusions}

We presented a strategy to optimize the parameters of the Omelyan integrator with $\alpha = 1/6$,
Hasenbusch mass preconditioning and three time-scales. Our method relies on the existence of a shadow Hamiltonian.\\
The schematic recipe followed by this work is the following:
\begin{itemize}
\setlength\itemsep{-0.5em}
\item Start with a \emph{reasonable} choice for the simulation of $(n, m, k)$. In the present work we used
conservative choices just to have a complete description to high values of $\mu$.
\item Measure the forces in each level, which we already computed for the evolution, and calculate $|\mathcal{F}_i|^2 = \sum_{x,\mu,a}T_{R,i} \left(F_i^{a\mu}(x)\right)^2$ and its variance ${\rm Var} (|\mathcal{F}_i|^2)$.
\item Measure the number of MVMs in each level.
\item By fitting the dependence in $\mu$ we are able to predict the acceptance and the cost dependence
on $(n, m, k, \mu)$ with accuracy within 10\%.
\end{itemize}
The minimization of the cost with this method is \emph{cheap} since it employs the forces already
calculated in the simulations. 
Generalizing it to a larger number of Hasenbusch levels on dif\mbox{}ferent quark determinant splitting \cite{Luscher:2005rx} is rather straightforward, especially as long as covariances can be neglected.
The results are encouraging and we plan to perform a study of the mass dependence 
as well as to consider dif\mbox{}ferent strongly interacting BSM models. Many aspects, not covered for space reason, will be discussed in a forthcoming publication.

\section*{Acknowledgments}

This work was supported by the Danish National Research Foundation DNRF:90 grant and by a Lundbeck Foundation Fellowship grant. The computing facilities were provided by the Danish Centre for Scientific Computing and the DeIC national HPC center at SDU.

\end{document}